\documentclass[a4paper]{jpconf}
\usepackage{graphicx}
\usepackage{lineno}
\usepackage{hyperref}
\usepackage{color}
\usepackage{url}

\begin{document}

\title{Heterogeneous High Throughput Scientific Computing with APM X-Gene and
Intel Xeon Phi}

\author{David Abdurachmanov$^1$, Brian Bockelman$^2$, Peter Elmer$^3$, Giulio
Eulisse$^4$, Robert Knight$^5$, Shahzad Muzaffar$^4$}

\address{$^1$ Digital Science and Computing Center, Faculty of Mathematics and
Informatics, Vilnius University, Vilnius, Lithuania}
\address{$^2$ University of Nebraska-Lincoln}
\address{$^3$ Department of Physics, Princeton University, Princeton, NJ 08540,
USA}
\address{$^4$ Fermilab, Batavia, IL 60510, USA}
\address{$^5$ Research Computing, Office of Information Technology, Princeton
University, Princeton, New Jersey 08540, USA}

\ead{David.Abdurachmanov@cern.ch}

\begin{abstract}
Electrical power requirements will be a constraint on the future
growth of Distributed High Throughput Computing (DHTC) 
as used by High Energy Physics. Performance-per-watt is a critical
metric for the evaluation of computer architectures for cost-efficient
computing. Additionally, future performance growth will come from
heterogeneous, many-core, and high computing density platforms with
specialized processors. In this paper, we examine the Intel Xeon
Phi Many Integrated Cores (MIC) co-processor and Applied Micro
\mbox{X-Gene} ARMv8 64-bit low-power server system-on-a-chip (SoC)
solutions for scientific computing applications.
We report our experience on software porting, performance and energy
efficiency and evaluate the potential for use of such technologies
in the context of distributed computing systems such as the
Worldwide LHC Computing Grid (WLCG).
\end{abstract}

\section{Introduction and Motivation}
Processing the data produced by High Energy Physics (HEP) experiments like
those at the Large
Hadron Collider (LHC)~\cite{LHCPAPER} at the European Laboratory for Particle
Physics (CERN)
requires significant
computing resources. The scale is beyond those available in
a typical single computer center. The Worldwide LHC Computing Grid (WLCG) was
established to provide the computing resources needed for the LHC
experiments, and used, for example, by the CMS and
ATLAS experiments for the discovery of the Higgs
boson~\cite{CMSHIGGS,ATLASHIGGS}. It is a distributed computing resource
across 170 computing centers in 40 countries. The CMS experiment,
for example, used between 80,000 and 100,000 x86\_64 cores
from the WLCG for its processing needs in 2012. Planned luminosity upgrades of
the LHC~\cite{HLLHC} will result in a 2-3 order of magnitude increase in dataset
sizes over the next 15 years, requiring commensurate increases in
processing capacity.

Intel is the leading company in general purpose server processors market.
Alternative solutions, like ARMv8 64-bit, aim to provide high-density and energy
efficient platforms for computing centers. These platforms delivered by multiple
vendors are optimized for specific market segments while ensuring compatibility by
running a common instruction set, i.e. a single ecosystem is shared. General
purpose solutions provide benefits to a full software stack unlike graphics 
processing units (GPU), which require software redesign and only applicable for 
portions of the application.

In this paper we focus on two general purpose central processors:
Applied Micro (APM) X-Gene 1 ARMv8 64-bit Server-on-Chip and Intel's
Xeon Phi coprocessor. We investigate these two platforms as
alternatives to the x86\_64 family of processors used in the WLCG today.
We base our results on performance (events per second) scalability
over power (watts) usage. The power measurements we provide in this
paper are for the silicon chip and not for a full computing node.

\subsection{Market}
According to Gartner, Intel sold 92 percent of all server processors
in 2013~\cite{PCMAGINTEL2013}.  
Distributed High Throughput Computing (DHTC) is also dominated by x86\_64.
In contrast, ARM Holdings is a leader in low power and high energy
efficiency processors market. Their business model differs from
Intel as they provide license agreements for their intellectual
property (IP) to partners.  Instruction set (ISA) licenses allow
partners to create custom silicon chips for wide variety of
applications: mobile phones, notebook, servers and many others. 
In addition to providing ISA licenses ARM
Holdings offers already designed and verified building blocks (CPU
and GPU cores, on-chip interconnect and similar) for manufacturing
of silicon chips solutions.  Each solution is designed to meet
specific market requirements, while keeping the compatibility by
running a common ISA, like ARMv8 64-bit. ARM Holdings and partners
introduced a new high energy efficient general purpose server
products in 2014.


Intel announced Xeon Phi platform for HPC market
in 2012. It is also known as Many Integrated Cores (MIC) computer
architecture. It is a many-core and long-vector machine combining
benefits of CPU and GPU into a single product. It is PCIe add-on
card and requires a host system Intel Xeon processors. It provides
high computational power and high energy-efficiency solution for
HPC market. 
We have previously
reported results for the Xeon Phi~\cite{CHEP13ARMPHI} CMS Software
(CMSSW) port (still incomplete due to issues with the Intel C++
Compiler).

We begin with a description of our efforts to port software and
validate ARMv8 64-bit, and then describe a power and computational
performance comparison between multiple architectures.

\section{Software Port and Validation of ARMv8 64-bit}
\subsection{APM X-Gene}
APM is the first company to introduce server grade and custom
designed ARMv8 64-bit Server-on-Chip solution, X-Gene 1 built on
40nm process by TSMC semiconductor foundry. A next generation X-Gene
2 built on 28nm processes is already being sampled. Princeton and
CERN were provided with XC-1 development boards with a single X-Gene
1 APM883208 silicon chip for software porting efforts. It is primarily
intended for speeding up development efforts.  Specification of
silicon chip is provided in table \ref{some_table}. We have been
working with APM for more than a year to deliver CMS Software (CMSSW)
and Open Science Grid (OSG)~\cite{OSG} software stack for ARMv8
64-bit.

\subsection{Software Porting Issues}
We have previously reported work to port the CMS Software (CMSSW)
to ARMv7 32-bit System-on-Chip silicon chips~\cite{ACAT13ARM,CHEP13ARMPHI}.
It allowed us to understand many issues arising when porting to
non-x86\_64 architectures and to prepare for ARMv8 64-bit platforms.
Indeed issues resolved for ARMv7 32-bit were often also relevant
for ARMv8 64-bit, however some issues specific to ARMv8 64-bit were
encountered:
\begin{itemize}
\item Update of \texttt{autoconf} package was required for successful
detection of ARMV8 64-bit platform, \texttt{aarch64}.
\item In cases where ARMv8 64-bit support was already present, software
packages were updated to the newer version.
\item We added ARMv8 64-bit support to the ROOT~\cite{bib:root} software
package. The ROOT (version 5) was extended  with \texttt{linuxarm64} target.
One important feature, Reflex dictionary generation, is not supported for ARMv8
64-bit. ROOT uses GCCXML as the underlying implementation for data reflection,
and GCCXML (based on GCC 4.2.1) is far too old for ARMv8 64-bit support.
However, because the data model -- LP64 -- is the same between x86\_64 and
ARMv8 64-bit, we were able (as a workaround) to pre-generate dictionaries on
x86\_64 and then compile them on ARMv8 64-bit. We found that such dictionaries
work successfully. More recent releases of ROOT (version 6) use LLVM/Clang in
place of GCCXML for dictionary generation, but this still does not support
ARMv8 64-bit due to the use of the old Just-in-Time (JIT) interface. This JIT
interface will be removed from LLVM/Clang after the version 3.5 release and
ROOT is scheduled to migrated to MCJIT interface in 2015~\cite{ARMV72007}.
Until this migration happens \texttt{rootcling} will not work on ARMv8 64-bit
systems.
\item Oracle Instant Client is only provided as proprietary binary blobs and
not as source code, and is not available for the ARMv7 32-bit and ARMv8 64-bit
architectures. However, none of the standard CMS workflows run on the WLCG
depend on Oracle. Only a few specialized applications used at CERN alone need 
Oracle, for example, loading detector calibrations.
\item Toolchain (\texttt{GCC} and \texttt{binutils}) issues were found and
reported upstream to the projects. The final \texttt{GCC} 4.9.1 and
\texttt{binutils} (\texttt{bfd} linker) 2.24 is now capable of compiling CMSSW.
However, one issue remains open with CERN Virtual File System (CVMFS) package
used for CMSSW distribution across Grid sites. We had to modify the way CVMFS 
components (\texttt{libcvmfs\_only.a}, \texttt{libz.a}, \texttt{libsqlite3.a},
\texttt{libcurl.a} and \texttt{libcares.a}) are merged into the final 
\texttt{libcvmfs.a} library.
\item \texttt{IgProf} memory and performance profiler ported to ARMv8 64-bit 
and \texttt{libunwind} library improved with fast backtracing capability for 
ARMv7 32-bit and ARMv8 64-bit~\cite{ACAT2014IGPROF}.
\end{itemize}

\begin{figure}[h]
\centering
\begin{minipage}{7.5cm}
\includegraphics[width=7.5cm]{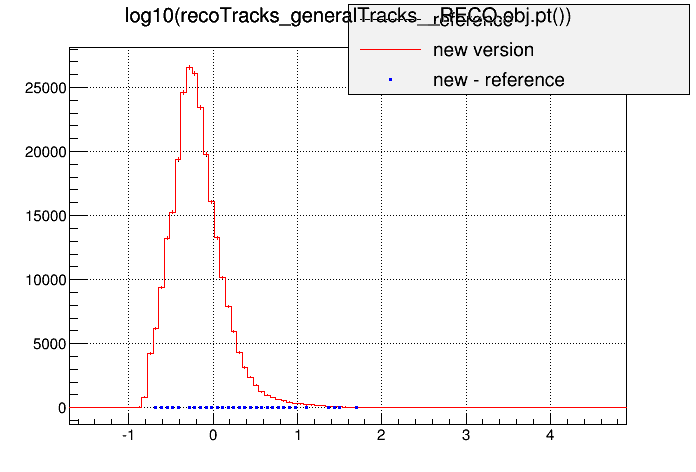}
\end{minipage}
\hspace{0.5cm}
\begin{minipage}{7.5cm}
\includegraphics[width=7.5cm]{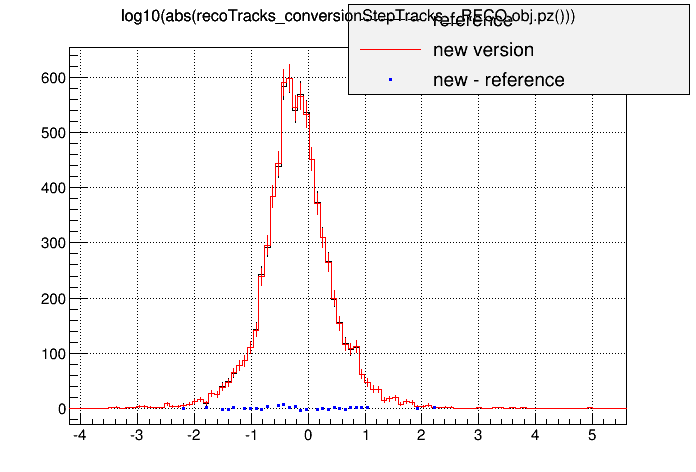}
\end{minipage}
\caption{\label{fig:recovalid}Example reconstruction validation distributions}
\end{figure}

\subsection{ARMv8 64-bit Validation}
To confirm that CMSSW provides comparable results on ARMv8 64-bit and
x86\_64, an initial validation was
done using the reconstruction software workflow. 
The
validation necessarily compared not only the two different architectures 
(ARMv8 64-bit to x86\_64), but also two different Linux 
distributions (Fedora 19 to Scientific Linux CERN 6.5).
Any discrepancies due to the latter, such as different
math libraries, should be resolved in
future validations. In addition, given that only a single board was 
available for ARMv8 64-bit, the validation was statistics limited 
to see any very subtle effects. 
With these minor caveats the observed differences are minimal 
(figure ~\ref{fig:recovalid} shows two examples) 
and within physics validation limits. 

\subsection{Grid Software}
A new architecture such as ARMv8 64-bit would likely be initially deployed
as a batch worker node. Thus we have also built the minimal set of required OSG 
software packages to support that: voms-clients,
HTCondor and CVMFS. CVMFS is used by CMS to distribute software and small 
data packages around Grid sites. Thus no local installation of these packages is 
required. CMSSW for ARMv8 64-bit is available alongside with CMSSW for x86\_64 
in CVMFS.
We have successfully used CMSSW release validation test suite on a local 
ARMv8 64-bit machine. It runs a number of predefined CMSSW (\texttt{cmsRun})
workflows with small number of events (10 or 100). The input for
workflows was generated locally as the first step or fetched from a
remote location. The CMSSW workflows successfully fetched
their input data from a remote \texttt{xrootd} server~\cite{xrootdpaper,xrootd-fed}, 
which demonstrates a functioning voms-clients package on ARMv8 64-bit.

We installed and configured HTCondor on XC-1 development board to
be a 8 slots (1-core and 2GB memory) worker node at Princeton
University. An additional x86\_64 machine was configured with
HTCondor as a master node for our ARMv8 64-bit worker node. The
master node was able to communicate with the worker node and successfully
execute the tasks.

\section{Test Environments for Power and Performance Measurements}
We now describe the test environments we have used to do power and performance
measurements for an Intel Xeon processor, an APM ARMv8 64-bit X-Gene 1
Server-on-Chip and
an Intel Xeon Phi coprocessor.
As described above, we have demonstrated standard CMSSW applications running
on the APM ARMv8 64-bit XC-1 board, however no three-way comparison of the
platforms
with CMSSW was possible due to the lack of a full CMSSW port on the Xeon Phi.
We have thus used the Geant4~\cite{GEANT4} benchmark, ParFullCMS,
as a simple cross-platform test capable of running in multi-threaded mode.
This benchmark uses a complex geometry (from CMS), but it is a
standalone application distributed with Geant4. Table
\ref{some_table} shows the selected general purpose platforms used for the
performance and energy efficiency benchmark.

\begin{table}[h]
\caption{\label{some_table}Silicon chips specifications}
\begin{center}
    \begin{tabular}{llll}
        \br
                             & X-Gene 1 & E5-2650 & Xeon Phi SE10/7120 \\
        \mr
            Physical cores   & 8 & 8 & 8 \\
            Threads per core & 1 & 2 & 4 \\
            Total threads    & 8 & 16 & 244 \\
            Frequency        & 2.4GHz & 2.0GHz & 1.24GHz \\
            Memory           & 16G (DDR3) & 256G (DDR3) & 16G (GDDR5) \\
        \br
    \end{tabular}
\end{center}
\end{table}

\subsection{Intel Xeon}
Our reference platform is a dual-socket Intel Xeon CPU E5-2650 @
2.00GHz launched in Q1 2012. It is an 8 physical core Sandy Bridge CPU
with hyper-threading (HT) enabled. The particular system was equipped
with 256GB DDR3 memory. For silicon-to-silicon comparison we only
measured power consumption of a single socket.
For the test, the ParFullCMS benchmark was compiled with
\texttt{GCC} 4.9.1.

Intel's Running Average Power Limit (RAPL)~\cite{INTEL64MAN,ACAT2014IGPROF}
technology was used for measuring power usage of different parts
of the silicon chip. It is available in modern Intel
micro-architectures starting from Sandy Bridge. It is
intended for controlling and limiting power usage in silicon chip.
In addition, RAPL provides capabilities to measure energy
and power usage. We used this capability to acquire readings from
the silicon chip.
The following RAPL domains are supported by server-grade silicon chips:
\begin{itemize}
    \item PP0 (Power Plane 0) -- processors cores subsystem.
    \item PKG (Package) -- processor die.
    \item DRAM (Memory) -- directly-attached DRAM.
\end{itemize}
The processors core subsystem consists of execution units, ALU,
FPU, L1 and L2 caches. The uncore subsystem (\texttt{energy(PKG) -
energy(PPO)}) consists of Intel QuickPath Interconnect (QPI), Last
Level Cache (LLc), on-chip memory and I/O. RAPL provides
$\sim$1ms~\cite{INTEL64MAN} resolution measurements. For the test
we only measured PKG power domain for a single socket. Reading RAPL
sensor data was done via model-specific registers (MSR) exposed to
Linux user-land via devfs.

\subsection{Intel Xeon Phi}
The Xeon Phi card is a highly parallel machine with 61 physical cores and 4-way
multi-threading running at 1.24GHz. It includes GPU-class 16GB GDDR5 memory, which
provide ultra-high bandwidth compared to standard DRAM solutions.
The card we used had only passive cooling. 
The card supports two execution modes: native and offloading. For
simplicity our tests were done using native execution on the card itself.
The Intel C++ Compiler
(Intel Composer XE 2013 SP1 Update 2, 14.0.2 20140120) was used to compile the
test as \texttt{GCC} does not provide Xeon Phi support.

The Xeon Phi card includes a number of sensors to acquire power usage
from different domains on the card. We used \texttt{libmicmgmt}
from Manycore Platform Software Stack (MPSS) 3.2 to read the sensor
data for different domains (PCIe, 2x3, 2x4, VCCP, VDDG, VDDQ, instant power and
more).
There is drawback of non-direct sensor reading. \texttt{libmicmgmt}
sends an interrupt over PCIe bus to uOS (Linux) forcing Xeon Phi
to switch to higher power state. This method cannot be used for
measuring idle power consumption. We measured idle power
consumption of the card by physically removing it from the server
enclosure. We found that idle power consumption was ~17W. The
resolution of sensors is 50ms. Two distinct datasets were used for
Xeon Phi: one for the whole card and another for VCCP + VDDG power
domains. According to Performance API (PAPI)~\cite{PAPI_MIC} source
code, VCCP power rail is processors core subsystem and VDDG --
uncore subsystem.


\subsection{APM X-Gene}
APM X-Gene 1 is 8 physical core processor running at 2.4GHz. The particular
system, XC-1 development board, contains 16GB DDR3 memory. The board provides
2 memory channels, but APM883408 is capable of addressing 512GB of memory using 4 memory channels.
Investigation done by APM showed that ParFullCMS test does not require high
memory bandwidth. We would like to emphasize that the firmware for managing
processor ACPI power states was not available in time for the test. We expect
X-Gene 1 efficiency to increase once the firmware is available. XC-1 is running
a custom build of Fedora 19 provided by Red Hat. The test was compiled with \texttt{GCC}
4.9.1. APM is working on X-Gene 1 optimizations for \texttt{GCC}, but changes were not
integrated upstream in time for the benchmark.

The XC-1 development board includes sensors for reading the power usage of
different domains of the silicon chip. We estimated power usage for  the silicon
chip by combining readings for PMD (core subsystem) and SOC (uncore subsystem)
power domains. The XC-1 is an IPMI-enabled device, thus readings from the sensors can
be acquired by \texttt{ipmitool} tool. We used IPMI before in previous
research, but we found that executing \texttt{ipmitool} is expensive and
restricts the measurements resolution. For XC-1 we read sensor data via I2C bus
and avoided
unnecessary overheads caused by execution of \texttt{ipmitool} tool.

\section{Results of Power and Performance Measurements}
\subsection{Platform Efficiency}
Our energy efficiency benchmark provided three-way silicon level
comparison between APM X-Gene 1 -- the first on the market ARMv8 64-bit
Server-on-Chip solution, Intel Xeon -- the dominating processor
family in server market and Intel Xeon Phi -- a highly parallel
long-vector processor family product intended for HPC workloads.
We sampled power usage readings from sensors with one second 
resolution on all platforms.

\begin{figure}[h]
\centering
\begin{minipage}{6.5cm}
\includegraphics[width=6.5cm]{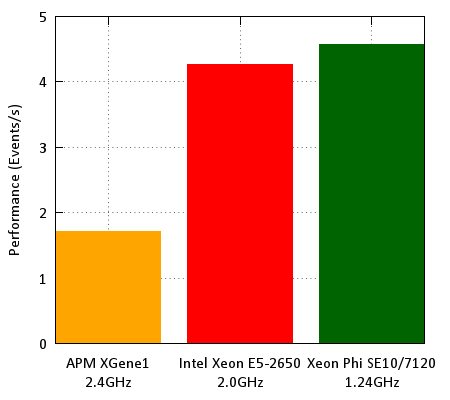}
\caption{\label{fig:platfomr_cmp}Absolute silicon chip performance}
\end{minipage}
\hspace{0.5cm}
\begin{minipage}{8.0cm}
\includegraphics[width=8.0cm]{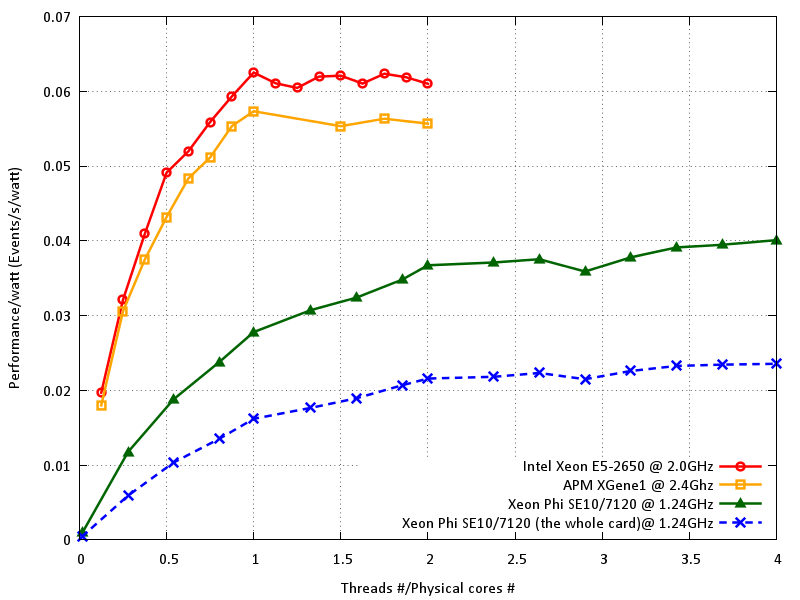}
\caption{\label{fig:power_scaling}Energy efficiency scalability}
\end{minipage}
\end{figure}

First we compared absolute performance provided by different solutions
by running on all available hardware threads (figure ~\ref{fig:platfomr_cmp}). 
We found that Intel Xeon Phi provided the
best performance, 1.07 times higher performance to Intel Xeon.
In the absence of the anticipated compiler optimizations, APM 
X-Gene 1 provided 2.48 times lower performance than Intel Xeon, 
but it was also operating at significantly lower power usage.

\begin{figure}[h]
\centering
\includegraphics[width=12cm]{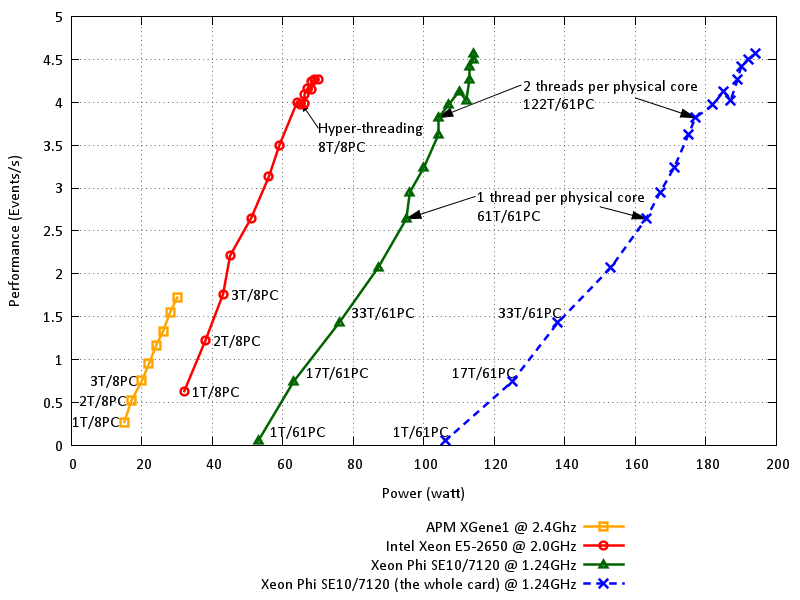}
\caption{\label{fig:some_comparison}Performance scalability over power usage}
\end{figure}

Then we measured how performance scales over power 
(figure ~\ref{fig:power_scaling}, figure ~\ref{fig:some_comparison}). We found
that APM X-Gene 1 running at full capacity (8 threads) was drawing
less power than Intel Xeon running a single thread and delivered
2.73 times higher performance. The Hyper-Threading (HT) on Intel
Xeon did not deliver higher energy efficiency. Additional few percent
of performance were gained at similar cost in power consumption
(figure ~\ref{fig:power_scaling}). We also over committed APM
X-Gene 1 with 2 threads per physical core, but we did not observe
significant change in energy efficiency. Intel Xeon Phi as
expected required at least 2 threads to achieve high utilization
of card.

\section{Conclusions}
We have built the software used by the CMS experiment at CERN, as well
as portions of the OSG software stack, for ARMv8 64-bit. 
It has been made available in the official CMS software
package repository and via the CVMFS distributed file system used by Grid
sites. Our initial validation has demonstrated that APM X-Gene 1 
Server-on-Chip ARMv8 64-bit solution is a relevant and potentially
interesting platform for heterogeneous high-density computing. In the absence 
of platform specific optimizations in the ARMv8 64-bit GCC compiler used, 
\mbox{APM X-Gene 1} shows excellent promise that the APM X-Gene hardware will 
be a valid competitor to Intel Xeon in term of power efficiency as the software 
evolves. However, Intel Xeon Phi is a completely different category of product.

As APM X-Gene 2 is being sampled right now, 
built on the TMSC 28nm process,
we look forward to extending our work to include it into our comparison.

\section*{Acknowledgements}
This work was partially supported by the National Science Foundation,
under Cooperative Agreement PHY-1120138, and by the U.S. Department
of Energy. We would like to express our gratitude to APM for providing
hardware and effort benchmarking Geant4 ParFullCMS, and to TechLab
at CERN for providing and managing the Intel Xeon and Intel Xeon Phi servers
for the benchmarks.

\section*{References}
\bibliographystyle{unsrt}
\bibliography{acat2014-armv8-xeon-phi}

\begin{thebibliography}{10}

\bibitem{LHCPAPER}
Lyndon Evans and Philip Bryant.
\newblock {LHC Machine}.
\newblock {\em JINST}, 3:S08001, 2008.

\bibitem{CMSHIGGS}
S.~Chatrchyan et~al.
\newblock {Observation of a new boson at a mass of 125 GeV with the CMS
  experiment at the LHC}.
\newblock {\em Phys.Lett.}, B716:30--61, 2012.

\bibitem{ATLASHIGGS}
G.~Aad et~al.
\newblock {Observation of a new particle in the search for the Standard Model
  Higgs boson with the ATLAS detector at the LHC}.
\newblock {\em Phys.Lett.}, B716:1--29, 2012.

\bibitem{HLLHC}
L~Rossi and O~Bruning.
\newblock {High Luminosity Large Hadron Collider A description for the European
  Strategy Preparatory Group}.
\newblock Technical Report CERN-ATS-2012-236, CERN, Geneva, Aug 2012.

\bibitem{PCMAGINTEL2013}
Michael~J. Miller.
\newblock Can {AMD}, {ARM}, or {IBM} dent {Intel}'s server dominance?
\newblock
  \url{http://forwardthinking.pcmag.com/none/323603-can-amd-arm-or-ibm-dent-intel-s-server-dominance},
  2014.
\newblock [Online; accessed 21-September-2014].

\bibitem{CHEP13ARMPHI}
David Abdurachmanov, Kapil Arya, Josh Bendavid, Tommaso Boccali, Gene
  Cooperman, Andrea Dotti, Peter Elmer, Giulio Eulisse, Francesco Giacomini,
  Christopher~D Jones, Matteo Manzali, and Shahzad Muzaffar.
\newblock Explorations of the viability of {ARM} and {Xeon Phi} for physics
  processing.
\newblock {\em Journal of Physics: Conference Series}, 513(5):052008, 2014.

\bibitem{OSG}
R~Pordes, D~Petravick, B~Kramer, D~Olson, M~Livny, A~Roy, P~Avery, K~Blackburn,
  T~Wenaus, F~Wuerthwein, I~Foster, R~Gardner, M~Wilde, A~Blatecky, J~McGee,
  and R~Quick.
\newblock The open science grid.
\newblock {\em Journal of Physics: Conference Series}, 78(1):012057, 2007.

\bibitem{ACAT13ARM}
David Abdurachmanov, Peter Elmer, Giulio Eulisse, and Shahzad Muzaffar.
\newblock Initial explorations of {ARM} processors for scientific computing.
\newblock {\em Journal of Physics: Conference Series}, 523(1):012009, 2014.

\bibitem{bib:root}
Rene Brun and Fons Rademakers.
\newblock Root — an object oriented data analysis framework.
\newblock {\em Nucl.Instrum.Meth.}, A389(1–2):81 -- 86, 1997.
\newblock New Computing Techniques in Physics Research V.

\bibitem{ARMV72007}
ARM Limited.
\newblock {ARM} discloses technical details and partner support for {ARMv7}
  architecture.
\newblock \url{http://www.arm.com/about/newsroom/8513.php}, 2005.
\newblock [Online; accessed 21-September-2014].

\bibitem{ACAT2014IGPROF}
David Abdurachmanov, Peter Elmer, Giulio Eulisse, Robert Knight, Tapio~Petteri
  Niemi, Jukka Nurminen, Filip Nyback, Goncalo~Marques Pestana, and Zhonghong
  Ou.
\newblock Techniques and tools for measuring energy efficiency of scientific
  software applications.
\newblock {\em To be published in ACAT 2014 proceedings}, 2014.

\bibitem{xrootdpaper}
A~Dorigo, P~Elmer, F~Furano, and A~Hanushevsky.
\newblock {XROOTD - A highly scalable architecture for data access}.
\newblock {\em WSEAS Transactions on Computers}, 4.3, 2005.

\bibitem{xrootd-fed}
L~Bauerdick, D~Benjamin, K~Bloom, B~Bockelman, D~Bradley, S~Dasu, M~Ernst,
  R~Gardner, A~Hanushevsky, H~Ito, D~Lesny, P~McGuigan, S~McKee, O~Rind,
  H~Severini, I~Sfiligoi, M~Tadel, I~Vukotic, S~Williams, F~Wuerthwein,
  A~Yagil, and W~Yang.
\newblock {Using Xrootd to Federate Regional Storage}.
\newblock {\em Journal of Physics: Conference Series}, 396(4):042009, 2012.

\bibitem{GEANT4}
S.~Agostinelli et~al.
\newblock {GEANT4: A Simulation toolkit}.
\newblock {\em Nucl.Instrum.Meth.}, A506:250--303, 2003.

\bibitem{INTEL64MAN}
Intel.
\newblock Intel 64 and {IA-32} architectures software developer’s manual.
  combined volumes: 1, {2A}, {2B}, {2C}, {3A}, {3B} and {3C}, September 2014.

\bibitem{PAPI_MIC}
\url{http://icl.cs.utk.edu/trac/papi/browser/src/components/micpower/README?rev=a7d7012}.

\end{thebibliography}

\end{document}